\documentstyle[prl,multicol,aps]{revtex}
\begin{document}
\input{epsf.tex} \epsfverbosetrue

\title{Observation of dipole-mode vector solitons}

\author{Wieslaw Krolikowski${}^{1}$, Elena A. Ostrovskaya${}^{2}$, Carsten Weilnau${}^{1,3}$, Matthias Geisser${}^{1}$, Glen McCarthy${}^{1}$, \\ Yuri S. Kivshar${}^{2}$, Cornelia Denz${}^{3}$, and Barry Luther-Davies${}^{1}$.}
\address{$^{1}$Laser Physics Centre, The Australian National University,
  Canberra ACT 0200, Australia}
\address{$^{2}$Optical Sciences Centre, The Australian National University,
  Canberra ACT 0200, Australia}
\address{$^{3}$Institute of Applied Physics,
Darmstadt University of Technology, Darmstadt, Germany}

\maketitle

\begin{abstract}{
 We report on the first experimental observation of a novel type of optical vector soliton, a {\em dipole-mode soliton}, recently predicted theoretically. We show that these vector solitons can be generated in a photorefractive medium employing two different processes: a phase imprinting, and a symmetry-breaking instability of a vortex-mode vector soliton. The experimental results display remarkable agreement with the theory, and confirm the robust nature of these {\em radially asymmetric} two-component solitary waves.}
\end{abstract}
\begin{multicols}{2}
\narrowtext

Optical spatial solitons in (2+1) dimensions are particle-like solitary waves propagating in a nonlinear bulk medium \cite{rev}. The exhaustive research of the past decade has shown that these ``light particles'' can possess topological phase properties analogous to a charge. Moreover, several light beams can combine to produce a {\em vector} soliton with a complex internal structure. This process can be thought of as the formation of a ``solitonic molecule'' from the constitutuents of different charge.  

Recently, the existence of the most robust ``solitonic molecule'', the {\em dipole-mode vector soliton}, has been predicted \cite{dipole}. This novel optical vector soliton originates from trapping of a dipole-mode beam by a waveguide created by a fundamental soliton in the co-propagating, incoherently coupled, beam. While many other topologically complex structures may be created, it is only the dipole mode that is expected to generate a family of {\em dynamically robust vector solitons} \cite{dipole}.  The closest counter-example is the {\em vortex-mode vector soliton} \cite{vortex} which has a node-less shape in one component and a ring-like vortex in the other component. This {\em radially symmetric},  vector soliton undergoes a nontrivial {\em symmetry-breaking instability} \cite{dipole,md}, which transforms it into a more stable object - {\em radially asymmetric} dipole-mode vector soliton, even in an isotropic nonlinear medium. 

While the existence and robustness of the dipole-mode vector solitons have been established theoretically for a general model of an isotropic medium with saturable nonlinearity \cite{dipole}, the main question still stands: {\em Is the stability of these asymmetric solitons, as opposed to the radially symmetric vortex-mode solitons, a fundamental phenomenon that can be demonstrated experimentally?} 

In this Letter we answer this question {\em positively}.
 We observe dipole-mode solitons in strontium barium niobate (SBN) photorefractive crystals experimentally, {\em by employing two different techniques}. First, we use a specially fabricated phase mask to create a dipole-like structure in one of the co-propagating, mutually incoherent, beams. Second, we observe the symmetry breaking of the vortex-mode soliton and the formation of a dipole-mode soliton, as predicted by the theory.

{\em Theoretical results.} We consider two {\em incoherently interacting optical beams} propagating in a  {\em bulk, isotropic, saturable medium}.  The model describes (2+1)-dimensional screening solitons in photorefractive (PR) materials in the isotropic approximation \cite{christ}. It represents a great simplification with respect to a more realistic treatment taking into account the inherent anisotropy of the nonlocal nonlinear response of a PR crystal \cite{Zozulya,Stepken}. However, we will show that {\em it does provide correct qualitative predictions} of the phenomena observed experimentally. Moreover, due to the generality of the model, one can expect to observe similar phenomena in other nonlinear isotropic or weakly anisotropic systems. 

The normalized equations for the slowly varying beam
envelopes, $E_1$ and $E_2$, can be written as follows \cite{christ,Buryak}: 
\begin{equation}\label{nls dyn}
i \frac{\partial E_{1,2}}{\partial z} + \Delta_{\perp} E_{1,2} - \frac{E_{1,2}}{ 1+ |E_{1}|^2 + |E_{2}|^2} = 0,
\end{equation}
where $\Delta_{\perp}$ is the transverse Laplacian, and $z$ is the propagation direction. Stationary, (2+1)-dimensional solutions of these equations can be found in the form
$E_1=\sqrt{\beta_1}\,u(x,y)\,\exp({i\beta_1z})$,
$E_2=\sqrt{\beta_1}\,w(x,y)\,\exp({i\beta_2z})$, where $\beta_1$ and $\beta_2$
are two independent propagation constants. Measuring the coordinates $x$ and $y$ 
in the units of $\sqrt{\beta_1}$, and introducing the soliton parameter $\lambda = (1-\beta_2)/(1-\beta_1)$, we derive the stationary equations for the normalized envelopes $u$ and $w$:
\begin{eqnarray}
\label{sat nls}
\Delta_{\perp} u - u + F(I)u= 0,\\ \nonumber
\Delta_{\perp} w - \lambda w +F(I)w = 0, 
\end{eqnarray}
where $F(I)=I(1+s I)^{-1}$, $I=u^2+w^2$, and $s=1-\beta_1$ is the saturation parameter. The limit $s \to 0$ corresponds to the Kerr medium \cite{md}.

As has been recently shown in \cite{dipole}, when one of the beams components, say $w$, is weak, a soliton formed by the $u$-component induces a change in the refractive index of the PR material that traps and guides the weaker $w$-component. If the waveguide is induced by a fundamental, bell-shaped soliton, its {\em nonlinear guided modes} form a set analogous to the {\em Hermite-Gaussian} and {\em Laguerre-Gaussian} {\em linear} modes supported by a radially symmetric waveguide \cite{Gagnon}. At higher intensities of the trapped beam, the two beams form a {\em vector soliton} that is self-trapped by a composite refractive index change induced by both beams. Apart from the bell-shaped vector solitons generated by the ground-state mode of the soliton-induced waveguide, one can anticipate various types of {\em higher-order (2+1)-dimensional vector solitons}. For example, a radially-symmetric {\em vortex-mode vector soliton} \cite{vortex,md} is generated by a node-less beam guiding the component with a nonzero topological charge and the doughnut structure of a Laguerre-Gaussian (${\rm LG}^1_0$) mode. A radially asymmetric {\em dipole-mode vector soliton} \cite{dipole}, has one node-less component, while the other component has a structure of a Hermite-Gaussian (${\rm HG}_{01}$) mode.
\begin{figure}
 \setlength{\epsfxsize}{8 cm} \centerline{ \epsfbox{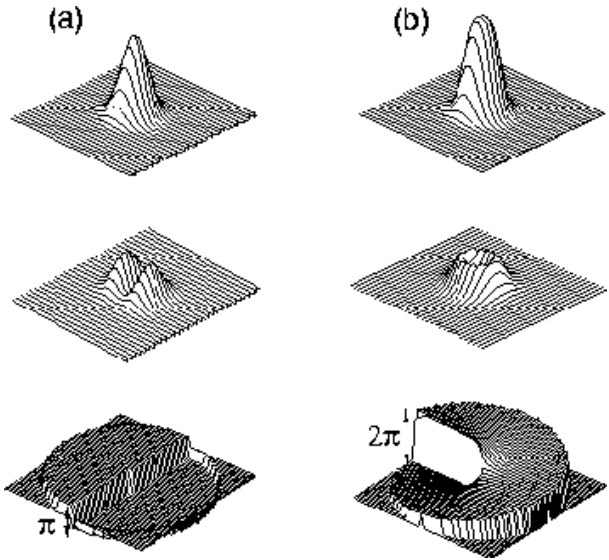}}
 \vspace{2mm}
  \caption{Examples of constituents of (a) dipole-mode and (b) vortex-mode vector solitons for $s=0.3$ and $s=0.65$, respectively (top and middle row) shown with the phase distributions needed to generate the $w$-component experimentally (bottom row).}
  \label{fig1}
\end{figure}
The existence domain for the radially asymmetric {\em dipole-mode vector solitons} of our model (\ref{sat nls}) was determined in Ref. \cite{dipole}. A typical example of such a soliton is shown in Fig. \ref{fig1}(a), along with a phase distribution in a dipole-mode constituent. For any given $s$, the solutions are characterized by a certain cut-off value $\lambda$, when the dipole-mode component vanishes. Near the cut-off, the vector soliton can be approximately described by the linear waveguiding theory. With increasing $\lambda$, the $w$ component grows and deforms the effective waveguide generated by the $u$-mode, so that the $u$-component elongates and becomes radially asymmetric. The {\em linear} and {\em dynamical stability analysis} of these solitons has revealed their astounding dynamical stability, with respect to both small- and large-amplitude perturbations \cite{dipole}. Overall, the analysis performed in \cite{dipole} suggested that the dipole-mode vortex solitons would lend themselves to experimental observation more easily than their unstable vortex-mode counterparts.

The family of radially symmetric {\em vortex-mode solutions} of Eqs. (\ref{sat nls}), in the form: $u=u(r)$, $w=w(r)\exp(im \phi)$, where $r=\sqrt{(x^2+y^2)}$, has been numerically found in \cite{dipole,md}. The $u$ component of this solution has no topological charge and the $w$ component carries a single-charged vortex ($m=\pm 1$).  In Fig. \ref{fig1}(b), we present a typical solution of this family, along with the helical phase distribution needed to generate a vortex in the $w$-component.  The {\em dynamical} and {\em linear stability analysis} of these solutions conducted in \cite{dipole,md} has proved that {\em all such vortex-mode vector solitons are linearly unstable}.  The instability growth rate is {\em positive} for any vanishingly small amplitude of the $w$ component, and it increases rapidly with the vortex intensity. 

Remarkably, an unstable vortex-mode soliton displays a {\em symmetry-breaking instability}. It {\em always decays into a radially asymmetric dipole-mode soliton} with nonzero angular momentum which can survive for {\em very large} propagation distances \cite{dipole}. The break-up of the vortex-mode vector soliton  in a saturable medium is a nontrivial effect which, as we show below, can be used to observe dipole-mode vector solitons in PR crystals.
\begin{figure}
  \setlength{\epsfxsize}{8.5cm} \centerline{ \epsfbox{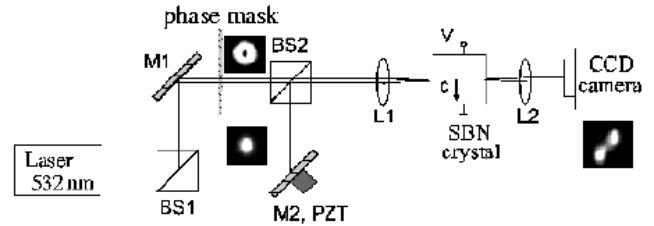}}
  \vspace{2mm}
  \caption{Experimental setup.}
  \label{fig2}
\end{figure}
{\em Experimental techniques.} Formation of dipole mode vector solitons  has been investigated using an
experimental setup shown schematically in Fig. \ref{fig2}.  An extraordinary polarized
laser beam  ($\lambda=532nm$) was split into two parts. First beam
was then transmitted through the phase mask (or glass slide) in order to
imprint the required phase structure.  In this way, we could obtain either an
optical vortex with intensity vanishing in the center of the beam [see Fig. \ref{fig1}(b)] or a 
dipole-like structure  with a phase jump of $\pi$ across the beam along
 its transverse direction [see Fig. \ref{fig1}(a)] that is perpendicular to the
optical axis of the crystal [the axis $(c)$ in Fig. \ref{fig2}].  The second beam was transmitted through
the  system of spherical or/and cylindrical lenses in order to form
either a circular or elliptically shaped spot of desired size.  The beams were
later combined using the beam splitter BS2 and focused into the input face
of the photorefractive crystal.
We used two samples  of the ferroelectric SBN crystal doped with Cerium (0.002\% by weight). Their dimensions $(a\times b \times c)$ are $(10\times 6
\times 5)$ $mm$ or $(15 \times 8 \times 5)$ $mm$. It is
well known that photorefractive crystals biased with strong DC electric
field exhibit strong positive or negative nonlinearity depending on the
polarity of the field \cite{screening}.  In our case, the SBN crystal was biased
with the DC field of $1.5-2.5$ $kV$ applied along  an optical axis
[the axis ($c$) in Fig. \ref{fig2}].  The resulting photorefractive nonlinearity was of  a self-focusing, saturable character. 
\begin{figure}
  \setlength{\epsfxsize}{8.0cm} \centerline{ \epsfbox{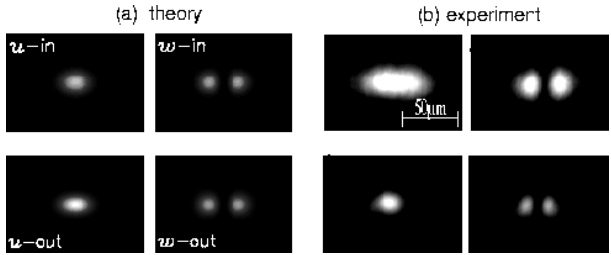}}
   \vspace{2mm}
  \caption{
    (a) Results of numerical simulations showing the generation of a stable dipole-mode soliton ($s=0.65$, $\lambda=0.5$, $z=35$). (b) Experimental results showing formation of a dipole-mode soliton (bottom) from the two out-of-phase beams (top, right) and co-propagating Gaussian beam (top, left). Experimental parameters are: $V=2$ $kV$, $z=15$ $mm$, and the initial powers of the $u$ and $w$ components $P_u=P_w=0.6$ $\mu W.$}
\label{fig3}
\end{figure}
\begin{figure}
  \setlength{\epsfxsize}{8.0 cm} \centerline{ \epsfbox{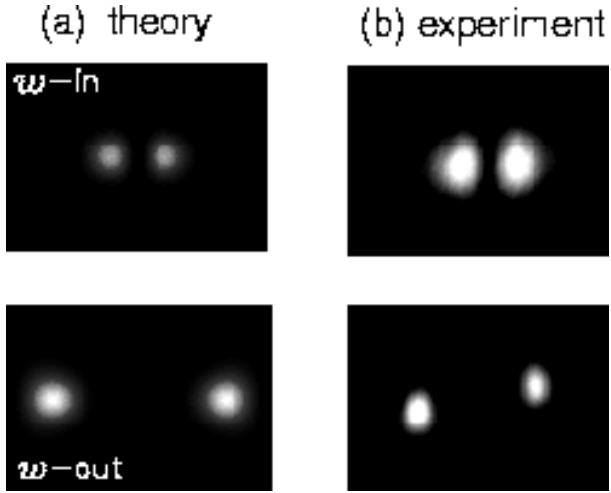}}
 \vspace{2mm}
  \caption{
    (a) Results of numerical simulations showing strong repulsion of the input dipole-mode lobes without the second component. (b) Experimental results showing the strong repulsion the two out-of-phase solitons formed by the dipole lobes without the co-propagating Gaussian beam. [Theoretical and experimental parameters as in Fig. \ref{fig3}(a,b)]. }
  \label{fig4}
\end{figure}
To
control the degree of saturation, we illuminated the crystal with a wide beam
derived from a white light source. We estimated the initial (i.e. at the input face of the crystal) degree of saturation to be of the order of unity in all our experiments.
Since both the components forming a vector soliton have to be mutually
incoherent, one of the constituent beams was reflected from a vibrating
mirror mounted on a piezoelectric transducer (M2, PZT). Driving this
transducer with an AC signal of $1$ $kHz$ imposed a frequency shift onto the
beam. This made both beams {\em effectively incoherent} inside the
photorefractive crystal as the slow photorefractive response does not follow fast changes of the relative phase between the beams. In all our
experiments the initial power of the beams did not exceed a few
microwatts.
\begin{figure}
  \setlength{\epsfxsize}{8 cm} \centerline{ \epsfbox{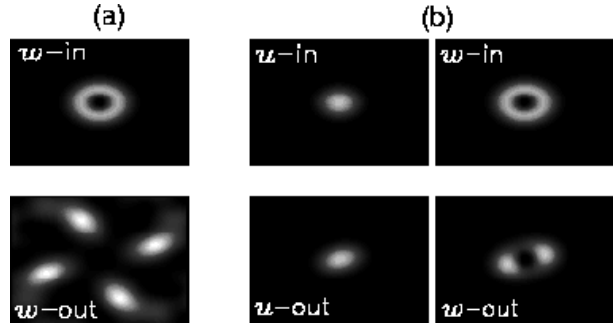}}
	\vspace{4mm}
  \caption{
   Suppression of the vortex filamentation due to modulational instability in the presence of a co-propagating fundamental-mode beam (see discussion in the text). Numerical results are obtained for $s=0.65$ and $\lambda=0.5$: (a) $z=35$, (b) $z=90$.}
  \label{fig5}
\end{figure}
\begin{figure}
  \setlength{\epsfxsize}{8cm}\centerline{\epsfbox{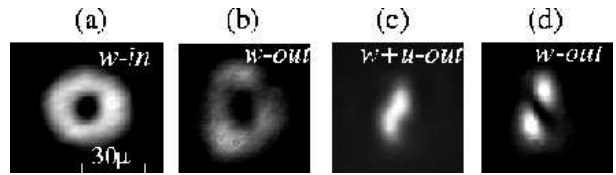}}
	\vspace{2mm}
  \caption{Experimental demonstration of suppression of the vortex filamentation in the presence of co-propagating fundamental-mode beam. Shown are (a) the vortex-bearing input beam, (b) output beam after propagating  without the Gaussian beam, (c) output intensity of the dipole-mode soliton formed by two co-propagating vortex and Gaussian beams, and (d) intensity of the dipole-mode constituent of the two-component vector soliton. Parameters are: $V=2.3$ $kV$, $P_u=P_w=0.3$ $\mu W$, $z=6$ $mm$.}
  \label{fig6}
\end{figure}

{\em Experiment vs theory.}  First, we generate the dipole-mode soliton by applying a phase mask to one of the input beams to create a dipole-like structure as in Fig. \ref{fig1}(a). If the dipole-mode bearing beam is launched {\em simultaneously} with a Gaussian beam shown in Fig. \ref{fig3}(b) (top row), and the two beams are made mutually incoherent, a robust dipole-mode vector soliton is generated [Fig. \ref{fig3}(b), bottom row]. However, when a dipole-mode-bearing beam is launched in the absence of a Gaussian beam, two out-of-phase lobes of the dipole beam form two coherently interacting fundamental solitons that {\em strongly repel each other} [Fig. \ref{fig4}(b)]. 

Our numerical simulations of the original equations (\ref{nls dyn}) provide excellent agreement with the experiment, as is seen in Figs. \ref{fig3}(a) and \ref{fig4}(a). The propagation distance of $z=35$ used in these simulations corresponds to $\approx 9$ $mm$ propagation distance in the PR crystal, which is {\em a priori} smaller than the crystal length in most experimental runs.

\begin{figure}
 \setlength{\epsfxsize}{8cm} \centerline{ \epsfbox{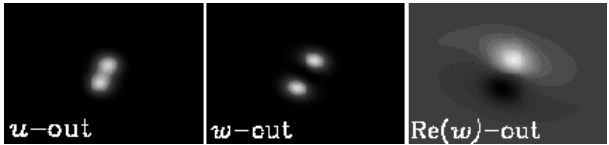}}
\vspace{2mm}
  \caption{Numerical results showing the intensity and phase distribution for the dipole-mode component of a vector soliton formed via decay of a vortex-mode vector soliton ($s=0.3$, $\lambda=0.6$, $z=30$).}
  \label{fig7}
\end{figure}

\begin{figure}
 \setlength{\epsfxsize}{5cm} \centerline{ \epsfbox{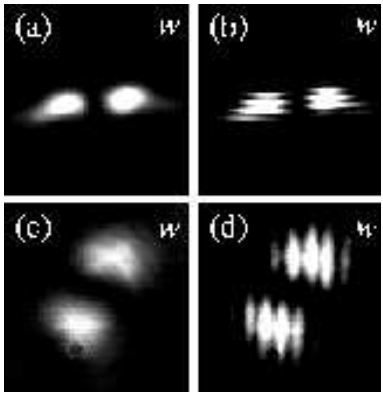}}
\vspace{2mm}
  \caption{Experimental results showing (a,c) the intensities  and (b,d) interference pattern for the dipole-mode lobes generated (a,b) via the phase-imprinting experiment, as in Fig. \ref{fig3}(b), and (c,d) via the symmetry-breaking of a vortex-mode soliton, as in Fig. \ref{fig6}(c,d). The shift seen in the interference fringes of the dipole-mode lobes overlapped with a mutually coherent plane wave, (b,d), is the result of the $\pi$ phase difference between the lobes. Parameters are: (a-b) $V=1.8$ $kV$, $P_u=P_w=1$ $\mu W$, $z=15$ $mm$; (c-d) corresponds to $V=1.8$ $kV$, $P_u=1.7$ $\mu W$, $P_w=0.4$ $\mu W$, $z=10$ $mm$.}
  \label{fig8}
\end{figure}

Next, we examine experimentally a non-trivial break-up of the vortex mode in the presence of the Gaussian beam.  Without the Gaussian beam, a scalar vortex exhibits filamentation due to the modulational instability \cite{fil_exp,Skryabin}. An intermediate stage of a complex vortex filamentation is shown in Fig. \ref{fig5}(a). The corresponding results of an experimental propagation of a single-charge vortex is shown in Figs. \ref{fig6}(a,b). The filamentation picture is different in the experimental situation owing to the  and the inherent anisotropy of the nonlocal nonlinear response of a PR crystal. 
When the vortex component $w$ is launched {\em simultaneously} with the uncharged component $u$, the scenario of the instability development {\em changes dramatically}. The vortex-mode instability leads to the formation of a dipole-mode vector soliton, as we demonstrate numerically [see  Fig. \ref{fig5}(b)] and experimentally [see  Figs. \ref{fig6}(c,d)]. Importantly, the phase distribution across the profile of the resulting dipole-mode beam calculated numerically and measured in the interference experiment [see Figs. \ref{fig7} and \ref{fig8}], confirms that the observed localized two-component structure is indeed the dipole-mode vector soliton predicted theoretically.
In conclusion, we have generated experimentally a novel type of vector optical soliton in a bulk medium. This soliton has a radially asymmetric structure and originates from trapping of a dipole mode by the soliton-induced waveguide, being much more robust than the corresponding vortex-mode vector soliton. Our experimental results are verified by a systematic comparison with the theory.

C. Weilnau's stay in Australia is supported by the Deutscher Akademisher Austauschdienst (DAAD). This work was supported by the Performance and Planning Fund of the Australian National University.

\end{multicols}
\end{document}